\documentclass[twocolumn]{aastex701}

\newcommand{\vunit}{\mbox{\,km\,s$^{-1}$}}

\newcommand{\Msun}{\mbox{\,$M_\odot$}}

\usepackage[normalem]{ulem}
\usepackage{color, soul}
\usepackage{graphicx}
\usepackage{epsfig}

\received{2025 Dec 14}
\revised{2026 Jan 19}
\accepted{2026 Jan 19}

\begin{document}

\title{The [\ion{Fe}{13}] Infrared 10\,747\AA\ and 10\,798\AA\ Lines in Novae}

\author[orcid=0000-0002-9670-4824,sname='Banerjee']{D. P. K. Banerjee}
\affiliation{Physical Research Laboratory, Navrangpura,  Ahmedabad, Gujarat 380009, India}
\email[show]{dpkb12345@gmail.com}  

\author[orcid=0000-0001-6567-627X,sname='Woodward']{C. E. Woodward}
\affiliation{Minnesota Institute for Astrophysics, School of
Physics \& Astronomy, 116 Church Street SE, University of Minnesota,
Minneapolis, MN 55455, USA}
\email[show]{chickw024@gmail.com}

\author[orcid=0000-0002-3142-8953, sname='Evans']{A. Evans} 
\affiliation{Astrophysics Research Centre, Lennard Jones
Laboratory, Keele University, Keele, Staffordshire,  ST5 5BG, UK}
\email{a.evans@keele.ac.uk}

\author[orcid=0000-0003-2824-3875,sname='Geballe']{T. R. Geballe}
\affiliation{Gemini Observatory/NSF's NOIRLab, 670 N. Aohoku Place, Hilo, HI, 96720, USA}
\email{tom.geballe@noirlab.edu}

\author[orcid=0000-0002-1457-4027,sname='Joshi']{V. Joshi}
\affiliation{Physical Research Laboratory, Navrangpura,  Ahmedabad, Gujarat 380009, India}
\email{onlyvishal@gmail.com}

\author[orcid=0000-0002-1359-6312,sname='Starrfield']{S. Starrfield}
\affiliation{School of Earth and Space Exploration, Arizona State University, Box 876004, Tempe, AZ 85287-6004, USA}
\email{sumner.starrfield@gmail.com}

\correspondingauthor{C.E. Woodward}

\begin{abstract}
The forbidden lines of [\ion{Fe}{13}] at 10\,747\AA\ and 10\,798\AA\  are among the most prominent lines in the
near-infrared spectrum of the solar corona. They have been used routinely, both outside and during eclipses, 
as sensitive probes of the electron density and polarization in the solar corona. Many novae pass through a 
coronal phase, wherein the highly ionized nova ejecta have physical conditions that are remarkably similar 
to those of the solar corona. Many  of the coronal emission lines that are seen are common to the spectra of both 
the Sun and novae. Yet, it appears that no robust detection of the [\ion{Fe}{13}] lines has been made in a nova. Here 
we report the detection of these two infrared [\ion{Fe}{13}] lines in the spectrum of the recurrent nova 
V3890~Sgr, taken 23.43 and 31.35~days after its August 2019 outburst. From their line strengths, we derive 
values of $10^{10}$~cm$^{-3}$ and $10^{[8.5-9]}$~cm$^{-3}$ for the electron density on the two
epochs. The decrease in density between epochs can be explained if the density decreased with a 
power law  $n(r)\propto{r}^{\alpha}$, with a $\alpha$ inferred to be $-3.$ The average temperature of the coronal 
gas is estimated to be T $= (2.51\pm0.06) \times 10^6$~K. We  find that recurrent novae with giant secondaries, 
including T~CrB whose eruption is imminent, are the most suitable sources for further detections of the [\ion{Fe}{13}] lines.
\end{abstract}


\section{Introduction} 

The presence of bright emission lines in the corona have been
observed during total solar eclipses since the 19th century,
the most prominent being the ``green line'' (\ion{Fe}{14} at
5303\AA)\footnote[1]{All wavelengths are in air.}, 
possibly first observed in 1869.  With the invention of the
coronagraph in the 1930s \citep{lyot39}, these lines could
be observed even outside of
eclipses. Lyot observed numerous coronal lines at Pic-du-Midi,
discovering the near-infrared 10\,747\AA\ and 10\,798\AA\
lines in 1936, for which he  employed  hyper-sensitization
techniques
to increase the sensitivity of photographic plates to infrared
radiation. The lines were eventually identified by
\cite{edlen45},
and shown to be forbidden transitions of \ion{Fe}{13}. The
lines are forbidden magnetic dipole (M1) transitions, and
come from very hot regions of the corona. 
The effective temperature at which \ion{Fe}{13} has its maximum
ionization fraction is T$_{\rm eff} = 1.8\times10^6$~K  
\citep[$\log{\rm{T}}_{\rm eff} = 6.25$;][]{schad23}. 
The ionization potential of \ion{Fe}{12} is 330.8~eV.

\begin{figure}[ht!]
\centering
\centering
\includegraphics[width=0.445\textwidth, trim={0.5cm 0.5cm 0.5cm 0.1cm}, clip, keepaspectratio]{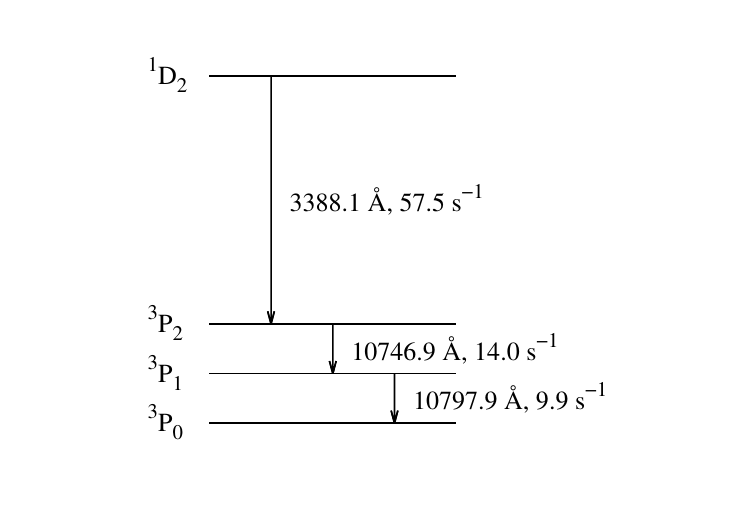}
\caption{Term diagram for the [\ion{Fe}{13}] lines of
interest. The transition probabilities are shown to the right of the wavelengths.
The $^3$P$_2$ and $^3$P$_1$ are 1.15~eV and
2.301~eV above the ground state respectively. 
Transition probabilities taken from NIST database
\url{https://physics.nist.gov/PhysRefData/ASD/lines\_form.html}
\label{term}}
\end{figure}

The two lines arise from different upper levels, as shown in
Figure~\ref{term}. They are a useful
diagnostic to determine the electron density and temperature,
as well as for magnetometry. In the latter context, there 
are a number of suitable forbidden lines in the solar corona
(e.g., [\ion{Fe}{15}] 5303\AA, [\ion{Fe}{10}] 6374\AA,
[\ion{Fe}{13} 10\,750\AA). Measurement of the full Stokes
vectors of these lines can give information on the magnetic
field properties. Indeed the magnetic sensitivity of the 
[\ion{Fe}{13}] 10\,747\AA\ line is 
significantly higher than lines in the visible, because of
Zeeman splitting, $\Delta\lambda\propto\lambda^2B$.

The 10\,747\AA\ line is  one of the strongest 
solar corona emission lines
in the near-infrared, and is widely used for solar
coronal analysis. However
in novae, the [\ion{Fe}{13}] lines have not been detected,
except perhaps in the extragalactic
recurrent nova M31-2008-12a, as discussed below.

Following their eruptions, classical and recurrent novae can reach a phase in which coronal lines
dominate the spectrum (``the coronal phase''). There are two possible sources for the high degree of
ionization needed to produce the coronal lines. One is photoionization by the central white dwarf (WD), 
whose observed surface temperature after eruption progressively increases
until the nova's pseudo-photosphere reaches the WD surface. During the ``super-soft source'' (SSS) X-ray 
phase, the WD surface temperature can reach peak values as high  as $100-110$~ev, ([$1.16-1.27]\times10^6$~K), 
and thus become a source of X-rays. The second mechanism for ionization, in the case when the secondary
star is a red giant (RG), is shock heating. Mass loss from the RG wind produces an extended envelope of 
material around the system. During the eruption, the nova's ejecta, traveling at typically  $\sim1000-3000$\vunit,
plows into the RG wind creating a strong forward shock and a reverse shock. The shock-heated material can 
reach temperatures as high as $10^7$~K \citep{banerjee14,drake16}. The nova ejecta, impeded by the wind, can 
decelerate sharply, and a decrease in velocity from 1000--3000\vunit\ to a few hundred \vunit\ is often observed in 
such systems \citep[e.g.,][]{das06,munari11, banerjee14,evans22}. The deceleration is inferred from the reduction 
of the emission line widths with time. Moreover, there can be strong shocks even without a RG wind, when
clumps of nova ejecta collide with each other. Such  ``internal shocks'' have been observed to occur
\citep[for example, in nova V959~Mon;][]{chomiuk14,chomiuk21}.

\citet[][henceforth AA23]{2023A&A...674A.139A} have carried out a multi-wavelength spectroscopic study of shock-driven phenomena 
in the outbursts of symbiotic-like recurrent novae (SyRNe), with an emphasis on RS~Ophiuchi. The 
systems in their study studied included RS~Oph itself, V745~Sco, V407~Cyg, and the subject of this
paper, V3890~Sgr. AA23 find that, during the eruptions of SyRNe, the line profiles exhibit
a broad component that originates in the ejecta and shock, and a narrow component that originates 
in the RG wind. With regard to the coronal emission, lines such as those of [\ion{Fe}{11}], [\ion{Fe}{12}], 
[\ion{Fe}{14}], may appear in post-maximum spectra of SyRNe eruptions. They are strongest in the 
earliest stages. These lines arise from two separate regions: the ionized wind of the RG (where 
T$_{\rm{e}} \sim 10^{4}$~K), and the post-shocked ejecta and wind (T$_{\rm{e}} > 10^{6}$~K);
the profiles of lines originating in the wind are narrow, while those of lines originating in the
post-shock gas are broad (AA23). The presence of [\ion{Fe}{10}] 
$\lambda6376$\AA, [\ion{Fe}{11}] $\lambda7982$\AA\ and [\ion{Fe}{14}] $\lambda5303$\AA\ lines in 
the SyRN V745~Sco is noted, but only the [\ion{Fe}{11}] $\lambda2648$\AA\ line is reported as being 
present in V3890~Sgr. We assume the infrared [\ion{Fe}{13}] $\lambda10\,747$\AA\ and
$\lambda10\,798$\AA\ lines (which were not discussed by AA23 as being
present in V3890~Sgr) originate in a similar (or nearby) zone as emission from the higher ionization [\ion{Fe}{14}] 
$\lambda$5303\AA\ line.

\subsection{Coronal lines in novae and the Sun}

Most of the coronal lines seen in novae are the same as those seen in the solar corona, thereby implying similar
physical conditions of temperature and density.  Even some of the most highly ionized species, that require 
photon energies of 285--350 eV to reach the energy levels from which lines such as [\ion{Fe}{14}] 5303\AA, 
[\ion{Al}{9}] 2.04\micron, and [\ion{Si}{9}] 1.43\micron\ in the infrared, are also routinely seen in novae. 
This implies that the nova ejecta must be very hot during the coronal stage. However, the [\ion{Fe}{13}] lines 
seen in the solar corona have not been identified in nova spectra, the likely reason being that they
are intrinsically much weaker than even the wings of the nearby \ion{He}{1} 10\,830\AA\ line on the redward 
side, and the strong \ion{C}{1} 10\,690\AA\ line on the blueward side \citep[the \ion{C}{1} line being especially 
strong in the so-called ``\ion{Fe}{2} novae'';][]{rudy00}. 

The \ion{He}{1} 10\,830\AA\ is the strongest line in the near-infrared, overwhelming all nearby
lines -- even the strongest H lines -- during the nebular phase. More importantly,  its wings can easily 
extend past the positions of the \ion{Fe}{13} lines, since the full width at zero intensity (FWZI) of the He line 
(or nova lines in general) can reach up to 10,000\vunit\ \citep[e.g.,][]{williams92}. In comparison, the separation
between the line centers of the \ion{He}{1} 10\,830\AA\ line and the more distant [\ion{Fe}{13}] 10\,757\AA\ line is 
only $\sim2030$\vunit. Thus the wings of the \ion{He}{1} line can engulf both of the nearby \ion{Fe}{13} lines. In contrast 
to this unfavorable situation in novae, in the solar corona the [\ion{Fe}{13}] and \ion{He}{1}  lines are  narrow and
distinctly resolved as separate lines \citep[e.g., Figure 3 of][]{schad23}. These lines have average full widths at half 
maximum (FWHM) $<50$\vunit, as is evident from the consideration that the FWHM of a line due
to thermal broadening is FWHM$_{\rm vel}= 0.2148\times(T/M)^{1/2}$\vunit, where $T$ is the gas temperature and 
$M$ is the ion mass in a.m.u. Thus, for \ion{Fe}{13}  and T$ = 10^{6}$~K, the FWHM$_{\rm vel} \simeq 29$\vunit\ for 
thermal broadening. Observed FWHMs in the solar corona are close to this value. 

\begin{figure}[h]
\centering
\vspace{0.1cm}
\includegraphics[width=0.485\textwidth, trim={2.0cm 3.5cm 3.0cm 2.4cm}, clip, keepaspectratio]{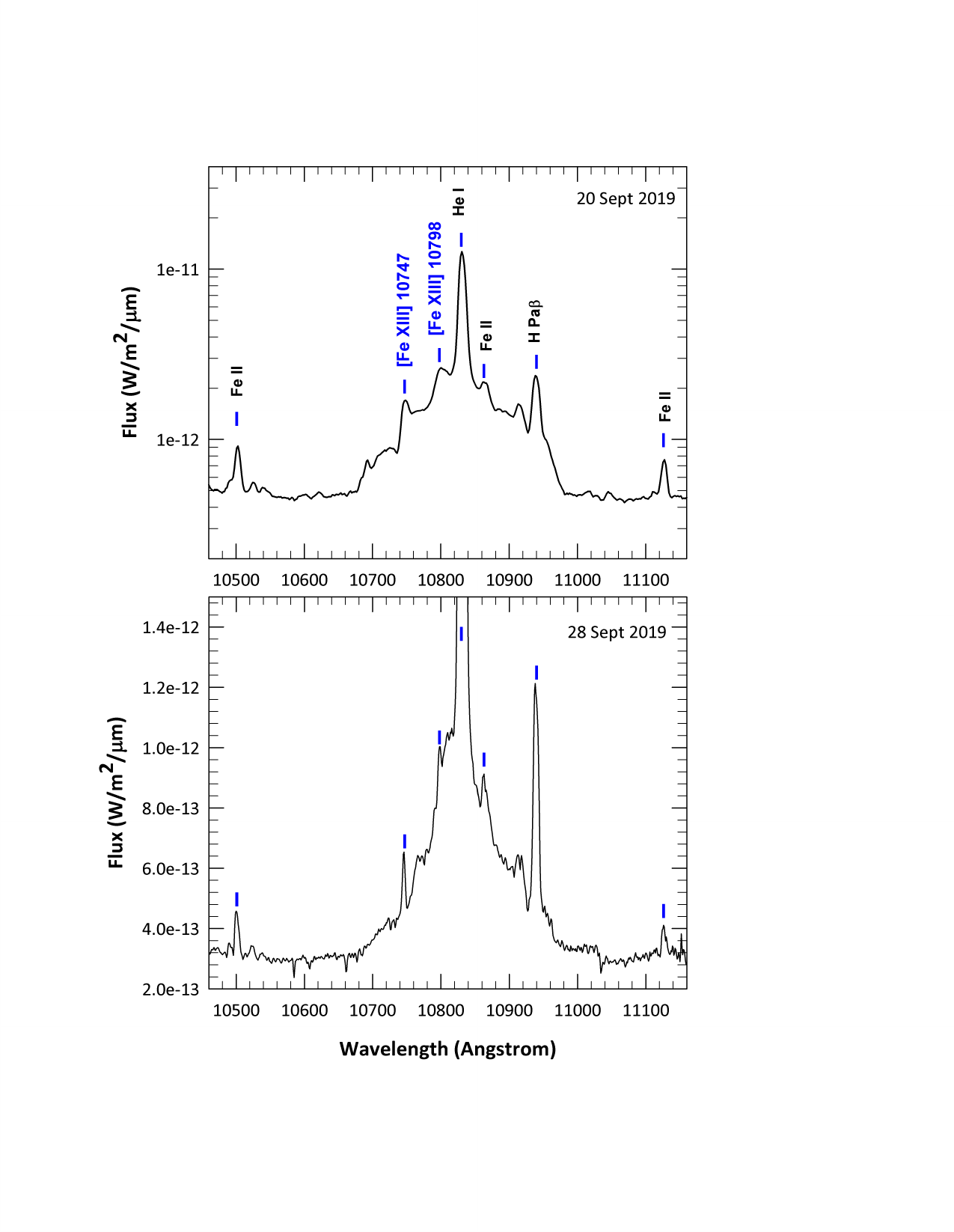}
\caption{The [\ion{Fe}{13}] 10\,747\AA, 10\,798\AA\ forbidden
lines, detected at two epochs, against the backdrop of the broad 
\ion{He}{1} 10\,830\AA\ line. The spectra also clearly reveal  
\ion{Fe}{2} 9\,997\AA\ (not shown here), 10\,501, 10\,863, and 
11\,126\AA\ --  the so-called  ``1 micron Fe II lines'' \citep{rudy00}. 
These descend from a common upper multiplet and are routinely detected
in novae and a variety of  emission-line stars. The blue ticks,
identifying the lines, are  marked at identical wavelengths in both
panels. The Y-axis in the bottom panel is on a linear scale to
emphasize the weakening of the  [\ion{Fe}{13}] 10\,798\AA\
line between epochs. 
\label{coronals}}
\end{figure}

However, if the secondary in the nova system is a RG and sufficient time has elapsed after eruption for the
emission lines to narrow to the order of 100\vunit, then the [\ion{Fe}{13}] lines may appear as distinct features 
in the spectrum. These are the conditions that prevailed when a [\ion{Fe}{13}] line was detected in the 
near-infrared spectrum of the extragalactic nova M31-N 2018-12a \citep{banerjee25}. This nova has a RG
secondary and its ejecta show a rapid deceleration after outburst. In view of its large distance,
the signal from M31-N 2018-12a was weak, even for an 8~m telescope. The \ion{Fe}{13} detection, which
we now believe was a blend of the two [\ion{Fe}{13}] lines, although not reported as such by \cite{banerjee25}, 
appeared genuine, but not so robust as to leave no room for doubt. However, in this paper we show convincing 
and unambiguous detections of the infrared [\ion{Fe}{13}] lines in (see Figure~\ref{coronals}) the recurrent nova V3890~Sgr.

\subsection{The 2019 eruption of V3890 Sgr}

Recurrent nova V3890 Sgr underwent eruptions in 1962 and 1990,
and had its latest eruption on 2019 Aug 27.87 
\citep[JD 2458723.37;][]{pereira19}. The system consists of a WD
and RG, with masses of 
$M_{\rm WD} = 1.35\pm0.13$\Msun\ and $M_{\rm RG} = 1.05\pm0.11$\Msun,
respectively \citep{schaefer09}. Detailed spectroscopic studies
of the nova in quiescence
and during its 2019 eruption are  given in \cite{kaminsky22} and 
\cite{evans22}, respectively. During the 2019 outburst,
several spectra were obtained by \cite{evans22} between 5.1 and 46.3
days after the eruption, and used to derive the elemental 
abundances and estimate the physical parameters and spatio-kinematics of the ejecta.

\section{Results}

\subsection{The Infrared [\ion{Fe}{13}] Lines}\label{sec:ObsFe}

We have analyzed the region around the infrared [\ion{Fe}{13}] lines in two specific spectra, from day 23.43 
(spectrum obtained by NASA IRTF /SpeX program 2020A-010) and 31.35 (spectrum obtained by
Gemini~N/GNIRS for program GN-2019B-DD-104). These are  shown in Figure~\ref{coronals}. Deceleration 
of the ejecta was
\clearpage 


\begin{longrotatetable}

\begin{deluxetable*}{ccccccccccccc}
\digitalasset
\tablewidth{0pt}
\tablecaption{Electron density and temperature from the [\ion{Fe}{13}] lines.
\label{nete}}
\tablehead{}
\startdata
UT Date & DAO$^a$ & Facility & Res & \multicolumn{3}{c}{Line flux (W~m$^{-2}$)$^{b}$} & $I$(10\,747\AA) & N$_{\rm{e}}$ (cm$^{-3}$)$^{c}$ & $I$(5303\AA) &
$\log(\rm{T})^d$& $I$(5303\AA) & $\log(\rm{T})^{e\!f}$ \\ \cline{8-8} \cline{10-10} \cline{12-12}
&&&&10\,747\AA&10\,798\AA&5303\AA&
$I$(10\,798\AA)
&&$I$(10\,747\AA)&&$I$(10\,798\AA)&\\ \hline
2019-09-20.30 &  23.43 & IRTF &  2000 &  $1.04(-15)$  & $8.88(-16)$ &  $4.39(-15)$ & 1.16--1.35 &  $1(10)$ &  4.23  & 6.41  &   4.94  &   6.40\\
2019-09-28.22  &  31.35 & GN &  6000  & $1.64(-16)$ &   $7.12(-17)$ & $3.95(-16)$ & 2.30 & $5(8)-1(9)$ & 2.41  & 6.38 &  5.55   &   6.39 \\
\enddata
\tablecomments{$^a$Days after outburst, with $t_0$ taken as 2019 Aug 27.87
(JD 2458723.37; Pereira 2019). $^{b}$Line fluxes are dereddened using 
$E(B-V) = 0.5$ from Evans et al. (2022).
$X(\pm{YY})$ denotes $X\times10^{\pm{YY}}$ for both line fluxes and
electron density.
$^{c}$Electron density estimated using tabular and other data
in Chevalier \& Lambert (1969) and  Malville (1967).
$^d\log(T)$ computed from the $I$(5303\AA/10\,747\AA) ratio  using  
Srivastava et al. (2007).
$^e{\log(T)}$ computed from the $I$(5303\AA/10\,798\AA) ratio  using  
Srivastava et al. (2007).
$^{f}$The mean temperature for the [\ion{Fe}{13}] lines on 
Sep 20 and 28 is $\log(\rm{T}) = 6.4 \pm 0.01$  (T$ = 2.51\times10^6$~K).
}
\end{deluxetable*}

\end{longrotatetable}

\noindent rapid in V3830~Sgr, and by days 23.43 and 31.35, the FWHMs of the lines,
deconvolved for instrumental broadening,  had decreased to 305\vunit\ and 230\vunit, respectively. In contrast, 
on Aug 28.895 (one day after the outburst) H$\alpha$, based on data from the 
ARAS\footnote[2]{https://aras-database.github.io/database/novae.html}
database \citep{teyssier19}, reveal that the H$\alpha$ line had a FWHM of $\sim5400$\vunit. 
The deceleration implies that a strong shock had developed in the system. Coronal lines 
developed early, those observed in the near-infrared and their evolution are given in \cite{evans22}. Based 
on the short duration of the SSS phase, which had ended by day 26, \cite{evans22} proposed that the
ions producing the coronal emission in the later spectra ($>23$~days) were solely collisionally ionized and excited. 

Figure~\ref{coronals} shows the principal lines seen, with their rest wavelengths in air denoted by blue 
tick-marks. The [\ion{Fe}{13}] lines are clearly seen at both epochs. Apart from \ion{He}{1} 10\,830\AA\ and 
hydrogen P$\gamma$, several ``1 micron \ion{Fe}{2} lines'' are also seen, some details of which are given 
in the caption of Figure~\ref{coronals}. The results of the analysis of the line ratios are presented
in Table~\ref{nete}. The observed line flux ratios, $I$(10\,747\AA)/$I$(10\,798\AA), listed in 
Table~\ref{nete}, were compared with theoretical predictions using tabular and graphical data from three 
independent studies, including those of  \cite{malville67}, \cite{chevalier69}, and \cite{dudik21}. 

\subsection{The Infrared [\ion{Fe}{13}] Lines and Electron Density}\label{sec:FeNe}

AA23 used the flux ratio
\begin{equation}
r = \frac{f(\mbox{\ion{Si}{3}] $\lambda1892$\AA})} {f(\mbox{\ion{C}{3}] $\lambda1909$\AA})}
 \end{equation}
\noindent obtained from low resolution IUE and Swift grism data, to estimate the electron density. Their 
Figure 15 suggests that this ratio is reasonably consistent over the 1990 and 2019 
eruptions.

\begin{figure}[h!]
\centering
\includegraphics[width=0.455\textwidth, trim={0.05cm 0.15cm 0.25cm 0.25cm}, clip, keepaspectratio]{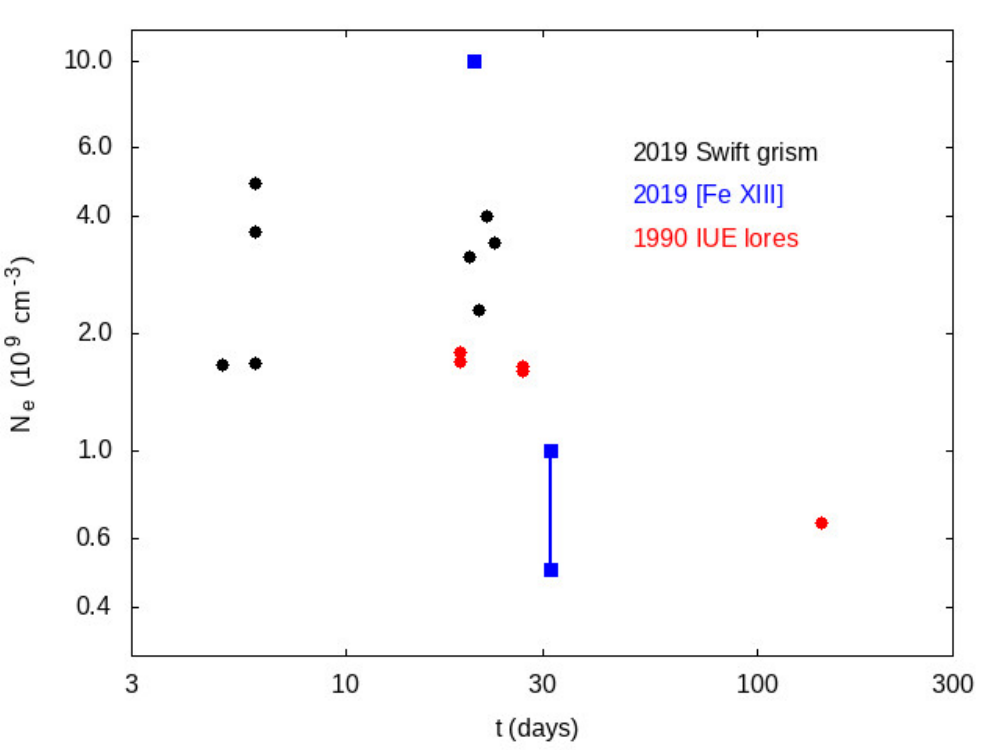}
\caption{Variation of the electron density (N$_{\rm{e}}$) for V3890 Sgr from \cite{2023A&A...674A.139A}; their 
values were binned and averaged over intervals of 2~days. The blue squares correspond to data from this infrared study.
See Table~\ref{nete} and text (section \ref{sec:FeNe}) for details.\label{Ne}
}
\end{figure}

Using the infrared [\ion{Fe}{13}] lines, we determine the electron density (N$_{\rm{e}}$) to 
be $\sim10^{10}$~cm$^{-3}$ on day 23.43 of the 2019 eruption, see Figure~\ref{Ne}. AA23 
obtained $1.74\times10^9$~cm$^{-3}$ and $1.63\times10^9$~cm$^{-3}$ respectively for days 19 and 27 
using low resolution IUE spectra (1990 outburst of V3890~Sgr), and $3.21\times10^9$~cm$^{-3}$
on day 21.5 from Swift grism data (2019 outburst).  Despite the large difference in the ionization states 
of the infrared lines used here and the IUE and Swift lines used by them to determine N$_{\rm{e}}$, 
our values are in reasonable agreement.

To determine the temperature, we used the analysis of  \cite{srivastava07}, which compares the 
ratio of the two [\ion{Fe}{13}] lines with the 5303\AA\ line.  For the 5303\AA\ flux,
we used flux-calibrated, high-resolution spectra on 2019 Sept 20.04UT and 26.014UT from the SMARTS 
database\footnote[3]{https://www.astro.sunysb.edu/fwalter/SMARTS/NovaAtlas/}
\citep{walter12}. These were the closest in time to our spectra
presented in Figure~\ref{coronals}. The four values of the gas
temperature, derived on both days and from both ratios
$I$(5\,303\AA)/$I$(10\,757\AA) and $I$(5\,303\AA)/$I$(10\,798\AA), 
are very similar and lie in the range $\log{\rm{T}} = 6.38-6.41$. These are 
indicated in Table~\ref{nete}; the  mean temperature of the four measurements 
is $\log{\rm{T}} = 6.40 \pm 0.01$~K (T $= [2.51\pm0.06]\times10^6$~K). In comparison, 
\cite{evans22}, using ratios of line fluxes from different sets of
coronal lines of sulphur and silicon, get an average value of 
$\log{\rm{T}} = 5.97\pm0.11$, or T$ = 9.3^{+2.7}_{-2.1}\times10^5$~K on day
23.43 (September 20.30). The highest temperature obtained by them for
that day is $\log{\rm{T}} = 6.17$ by using lines of [\ion{S}{9}] 
and [\ion{S}{12}]. The derived temperature here is in reasonable
agreement with values in \cite{evans22}.

It is puzzling why the density dropped by a factor exceeding 10 between day $\sim23$ to $\sim31$.
(Table~\ref{nete} shows a change from $10^{10}$~cm$^{-3}$ to $10^{8.5-9}$~cm$^{-3}$ in the $\sim8$ 
day period between day 23.43 and 31.35). The drop is not an artifact of the analysis. Visual 
examination of Figure~\ref{coronals} shows convincingly how much weaker the 10\,798\AA\ line has become in 
the second epoch, boosting $I$(10\,747\AA)/$I$(10\,798\AA) by almost a factor of two between epochs. 
For the Sun, a decrease in density with distance from the limb leads to higher values of $I$(10\,747\AA)/$I$(10\,798\AA). 
As examples, see Figures~2 of \cite{malville67} and \cite{dudik21}, and Table~IV of \cite{chevalier69}. However, 
if a decrease in the density of the ejecta occurred due solely to its expansion at constant velocity, it would be 
by a factor of $1.8 = (31.35/23.43)^2$ between the 2 epochs if the dilution is geometric 
(N$_{\rm{e}} \propto r^{-2}$, or equivalently  N$_{\rm{e}} \propto{t}^{-2}$).  If the dilution is steeper 
\citep[e.g., N$_{\rm{e}} \propto{r}^{-3}$;][]{1992ApJ...393..307H, 1997ApJ...490..803H}, 
then the density would have decreased by a factor of 2.4. But this still does not explain the observed 
data which shows a larger decrease in density. Indeed, the observed decrease in N$_{\rm{e}}$ requires 
that the decline must be at least as steep as $t^{-8}$, which seems unreasonable.

\begin{figure}[ht!]
\centering
\includegraphics[width=0.485\textwidth, trim={2.0cm 4.5cm 2.0cm 1.5cm}, clip, keepaspectratio]{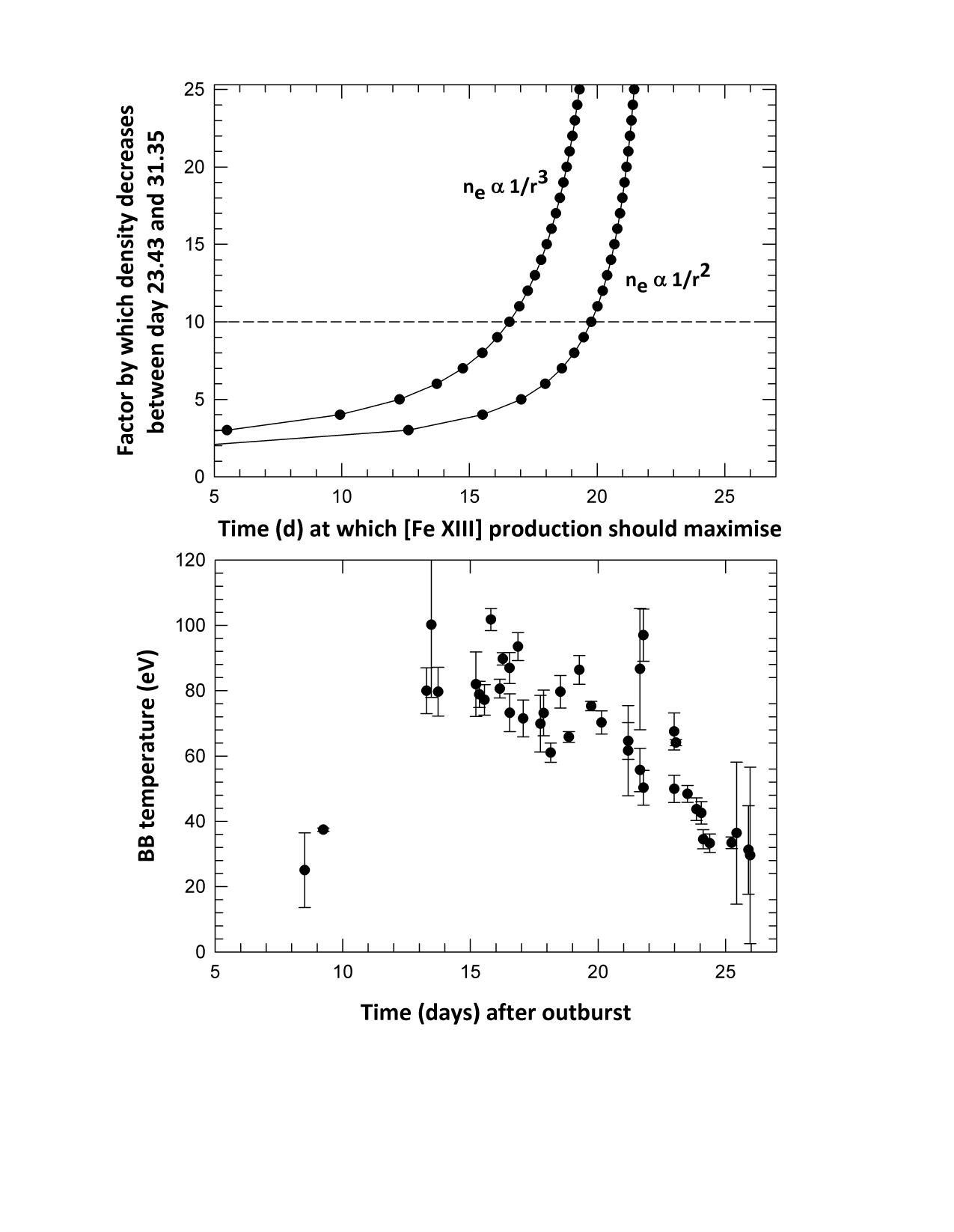}
\caption{The continuous lines show the time at which the bulk
of the [\ion{Fe}{13}] should be produced to explain the  drop
in density of the ejecta by a factor of 10 or more in 8 days.
The dashed line gives  an example that a drop in density by 10,
requires that the [\ion{Fe}{13}] ion production  maximize at
around day 16 (for N$_{\rm{e}} \propto{r}^{-3}$ expansion law).
 The bottom panel, based on Swift observations, shows the evolution 
of the black body temperature of the white dwarf during the SSS phase (see Section~\ref{sec:FeNe},
as well as  \citet{page20}).
\label{density_drop}}
\end{figure}

An alternative explanation for this behavior could be the following. We suppose that the [\ion{Fe}{13}] production
maximized on some day $t_{\rm XIII} ~~ (<23.43~\mbox{days})$  after outburst. If the dilution of the ejecta 
follows the form N$_{\rm{e}} \propto{r}^{-n}$, then to account for an increase by a factor  $\alpha\simeq10-25$
in the $I$(10\,757\AA)/$I$(10\,798\AA) ratio between days 23.43 to day 31.35, the following equation should hold
\begin{equation}
 [(31.35 – t_{\rm XIII})/(23.43 - t_{\rm XIII})]^n = \alpha\:\:.
\end{equation}
This is solved for $t_{\rm XIII}$ and  the solution is plotted in Figure~\ref{density_drop} for values of $n = 2, 3$.  
It is seen that for $n=3$, the density will decrease by a factor of 10 (as is close to what is  observed) if most of 
the [\ion{Fe}{13}] ions are produced on day 16.65.  The blackbody temperature, derived from the 
super soft X-ray emission,  also peaked near day 16 at $\sim 101.8$ ev (see Figure~\ref{density_drop} (bottom))
after the onset of the SSS phase.  A value of  $n=3$  is frequently used in CLOUDY \citep[e.g.,][]{2025arXiv250801102G} 
modeling, and may correspond to a homologous flow \citep{1992ApJ...393..307H, 1997ApJ...490..803H}. 

\begin{figure}[ht!]
\centering
\includegraphics[width=0.485\textwidth, trim={2.0cm 8.5cm 6.0cm 8.5cm}, clip, keepaspectratio]{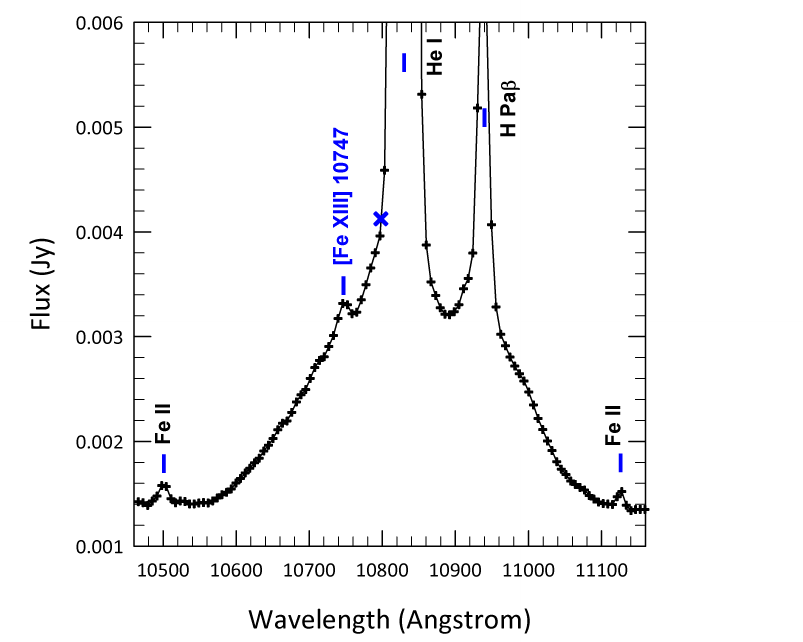}
\caption{The [\ion{Fe}{13}] 10\,747\AA\ line in SN 1987A.
This figure was prepared using JWST/NIRSPEC data available from the 
Mikulski Archive (MAST). The original study is described in 
\cite{larsson23}, who found several high-ionization coronal lines
from the equatorial ring of SN~1987A, requiring a temperature
$\ge2\times10^6$~K. As in Figure~\ref{coronals}, we identify two
of the ``1 micron Fe lines'' here to show the similarity with the
profile in V3890~Sgr.  A blue "X" shows the expected position
of the 10\,798\AA\ Fe line, which however is not seen. 
The observed spectrum had to be shifted blueward by 13\AA\ to
make the H, He and \ion{Fe}{2} spectral features match with
their rest wavelengths.
\label{87A}}
\end{figure}

\section{Discussion and conclusion}

Apart from the 10\,747\AA\ and 10\,798\AA\ lines, it is also possible to search in novae for the 
3388\AA\ line (see the term diagram in Figure~\ref{term}). However, detection of that line
should be challenging; it is accessible from the ground from a few observing sites,
but also it is intrinsically weak, about one-fifth of the 10\,757\AA\ line in intensity \citep{chevalier69}. It 
is a useful alternative as a density diagnostic, as can be seen from  the data in  \cite{chevalier69}, where 
the predicted ratios of 3388\AA\ relative to [\ion{Fe}{13}] infrared lines are listed over a
range of densities and temperatures. The line can be covered, for example, by the ultraviolet and visible 
grisms onboard a facility such as Swift, and can be observed from the ground, for example from
the Sutherland site of the South African Astronomical Observatory
\citep{nordsieck01}.

To obtain supplementary support for our detections of the  
[\ion{Fe}{13}] lines here, we have searched for detections  by others.
It transpires that there are very few instances where the
[\ion{Fe}{13}] lines have been detected. Among them, the
10\,747\AA\ line was reported in the symbiotic star R~Aqr
\citep{zirin76}, but the image was recorded on photographic
plates and is not available for inspection. The 10\,798\AA\
line is neither seen nor marked on the photographic plate. 

This study shows that the likelihood of making more detections of the
[\ion{Fe}{13}] lines in other novae will be higher under the following
conditions: (a)~the nova should have a  giant secondary with a wind to
decelerate the nova ejecta leading to the narrowing of emission
lines, which could occur in a recurrent nova or even a symbiotic
nova \citep[e.g., V407 Cyg;][]{banerjee14} with a RG/Mira secondary;
(b)~the nova should be in the coronal stage; (c) the lines should have
narrowed sufficiently that the  [\ion{Fe}{13}]  lines stand out
sharply and distinctly even if they are on the wings of a 
much-stronger  \ion{He}{1} 10\,830\AA\ line;  
and (d)~intermediate ($R > 2000$) or higher resolution is available
to allow the lines to stand out and thus facilitate detection.

In this context, the recurrent nova T~CrB, which has a RG secondary,
seems to be a suitable and attractive candidate. T~CrB is  expected to
erupt imminently. Its outburst is keenly awaited by the astronomical
community because it erupts only once   every 80 years -- its last
outburst was in 1946. It has a mysterious second maximum in the
visual light curve, and also displays an uncommon phenomenon of
its spectrum  switching from the He/N to \ion{Fe}{2} class during
its evolution \citep[see][]{morgan47}. Since the infrared [\ion{Fe}{13}]
lines are very sensitive density indicators,
they could give useful information on  T~CrB (and other systems) if these lines are
detected in its spectrum. 

\begin{acknowledgments} 
The authors wish to thanks the prompt and insightful comments
of the referee that improved this manuscript.

The IRTF data presented in this paper were obtained partly
under IRTF program 2020A-010. The Infrared Telescope Facility
is operated by the University of Hawaii under contract 80HGTR19D0030
with the National Aeronautics and Space Administration.
The Gemini observations were made possible by awards of
Director's Discretionary Time for program GN-2019B-DD-104.
The international Gemini Observatory is a program of 
NSF's NOIRLab, which is managed by the Association of Universities 
for Research in Astronomy (AURA) under a cooperative agreement 
with the National Science Foundation, on behalf of the Gemini 
Observatory partnership: the National Science Foundation 
(United States), National Research Council (Canada), 
Agencia Nacional de Investigaci\'{o}n y Desarrollo (Chile), 
Ministerio de Ciencia, Tecnolog\'{i}a e Innovaci\'{o}n (Argentina), 
Minist\'{e}rio da Ci\^{e}ncia, Tecnologia, 
Inova\c{c}\~{o}es e Comunica\c{c}\~{o}es (Brazil), 
and Korea Astronomy and Space Science Institute (Republic of Korea).

This work is based in part on observations made with the NASA/ESA/CSA
James Webb Space Telescope. The data were obtained from the Mikulski
Archive for Space Telescopes at the Space Telescope Science Institute,
which is operated by the Association of Universities for Research in
Astronomy, Inc., under NASA contract NAS 5-03127 for JWST. These
observations are associated with program \#1726 and can be accessed
via doi 10.17909/175h-7x33. 

SS acknowledges partial support from a NASA Emerging Worlds grant to ASU (80NSSC22K0361)
as well as support from his ASU Regents’ Professorship. CEW acknowledges partial
support from NASA grant JWST-GO-01731.006-A. We acknowledge the use of the SMARTS 
database and also thank Umberto Sollecchia who obtained the ARAS spectra used here.
\end{acknowledgments}

\facilities{Gemini:Gillett (GNIRS),  IRTF (SpeX), JWST}

\clearpage

\bibliography{bib_FeXIII_v7}{}
\bibliographystyle{aasjournalv7}

\end{document}